\documentclass[aps,epsfig,prl,twocolumn,superscriptaddress,showpacs,amssymb,nobibnotes]{revtex4}
\usepackage{graphicx,amsmath,amssymb}
\begin{document}
\title{Fluctuations and oscillations in a simple epidemic model}
\author{G. Rozhnova}
\author{A. Nunes}
 \affiliation{Centro de F{\'\i}sica Te{\'o}rica e Computacional and
Departamento de F{\'\i}sica, Faculdade de Ci{\^e}ncias da Universidade de
Lisboa, P-1649-003 Lisboa Codex, Portugal}
\begin{abstract}
We show that the simplest stochastic epidemiological models
with spatial correlations exhibit two types of oscillatory behaviour
in the endemic phase. In a large parameter
range, the oscillations are due to resonant amplification of stochastic
fluctuations, a general mechanism first reported for
predator-prey dynamics. In a narrow range of parameters that
includes many infectious diseases which confer long lasting immunity
the oscillations persist for infinite populations.
This effect is apparent in simulations of the stochastic process in systems
of variable size, and can be understood from the phase diagram of the
deterministic
pair approximation equations. The two mechanisms combined play a central
role in explaining the ubiquity of oscillatory behaviour in real data and in simulation results
of epidemic and other related models.
\end{abstract}
\maketitle
Cycles are a very striking behaviour of prey-predator systems also seen in a variety of other host-enemy systems {\textemdash} a case in point is the pattern of recurrent epidemics of many endemic infectious diseases {\cite{epid_osc}}. The controversy in the literature over the driving mechanisms of the pervasive noisy oscillations observed in these systems has been going on for long {\cite{ref_contro}}, because the simplest deterministic models predict damped, instead of sustained, oscillations. One of the aspects of this controversy is whether these mechanisms are mainly external or intrinsic, and the effects of seasonal forcing terms {\cite{seasonal}}, {\cite{bauschearn}} and of higher order non-linear interaction terms {\cite{mathlit}} have been explored in the framework of a purely deterministic description of well-mixed, infinite populations. These more elaborate models exhibit oscillatory steady states in certain parameter ranges, and have led to successful modelling when external periodic forcing is of paramount importance {\cite{grenfell_measles}}, but they fail to explain the widespread non-seasonal recurrent outbreaks found, for instance, in childhood infectious diseases {\cite{bausch}}.

During the last decade, important contributions have
come from studies that highlight the inherently stochastic nature of population dynamics
and the interaction patterns of the population as  important endogenous factors of recurrence
or periodicity {\cite{physics_sims}}.
A general mechanism of resonant amplification of demographic stochasticity has been 
proposed to describe the cycling behaviour of prey-predator systems {\cite{McKane-Newman2005}} and applied recently to recurrent epidemics of childhood infectious diseases {\cite{McKane_inter}}. 
The role of demographic stochasticity modelled as 
additive Gaussian white noise of arbitrary amplitude in sustaining incidence
oscillations had long been acknowledged in the literature 
{\cite{bartlett}}. The novelty in {\cite{McKane-Newman2005}} and {\cite{McKane_inter}}
was that of providing an analytical description of demographic stochasticity as an internal
noise term whose amplitude is determined by the parameters and the size of the 
system using a method originally proposed by van Kampen {\cite{vanKampen}}.

Our goal is to extend this approach  
by relaxing the homogeneous mixing assumption 
to include an implicit representation of spatial dependence. 
We show that the inclusion of correlations at the level of pairs leads to different quantitative and qualitative behaviours in a region of parameters that corresponds to infectious diseases which confer long lasting immunity. 
Our motivation was twofold.
On one hand, the homogeneous mixing assumption 
is known to give poor results for lattice or network structured population {\cite{lebowitz}},
{\cite{rev_keeling}}.
On the other hand, systematic simulations of infection on small-world networks  
have shown that the resonant amplification of stochastic fluctuations is significantly enhanced in the presence of spatial correlations {\cite{interface}}. Therefore, apart from stochasticity, the correlations due to the contact structure 
are another key ingredient to understand  the patterns of recurrent epidemics.
One of the main difficulties in including this ingredient is that the relevant network
of contacts for disease propagation is not well known 
{\cite{rev_keeling}}.
In this paper we shall consider a stochastic Susceptible-Infective-Recovered-Susceptible (SIRS)
epidemic model that leads 
to the ordinary pair approximation (PA)
equations of {\cite{lebowitz}} in the thermodynamic limit
as the simplest representation of the spatial correlations on an arbitrary network of fixed
coordination number $k$. The power spectrum of the fluctuations around the steady
state can be computed following the approach of 
{\cite{McKane-Newman2005}} and {\cite{vanKampen}}.
The combined effect of stochasticity and spatial correlations has been much studied through
simulations, but this is an analytical treatment of a model that includes
both these ingredients. 

\begin{figure}
{\includegraphics[width=\columnwidth]{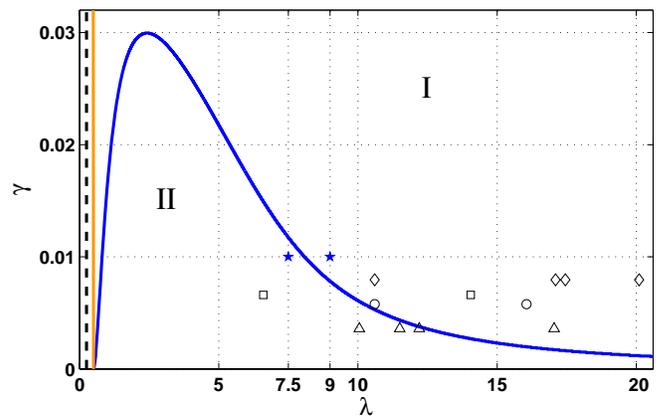} \caption{(Color online) Phase diagram in the $(\lambda,\gamma)$ plane for the MFA and the PA deterministic models
and parameter values for measles ($\vartriangle$), chicken pox ($\circ$), rubella ({\tiny $\square$}) and pertussis ($\diamond$)
from data sources for the pre-vaccination period. The blue stars are the parameter values
used in Fig. \ref{figure3}.} \label{phasediagramSIRS}} 
\end{figure}

Consider then a closed population of size $N$ at a given time $t$, consisting of $n_1$ individuals of type $S$, $n_2$ individuals of type $I$, and $(N-n_1-n_2)$ individuals of type $R$, modelled as network of fixed coordination number $k$. Recovered individuals lose immunity at rate $\gamma $, infected individuals recover at rate $\delta $, and infection of the susceptible node in a susceptible-infected link occurs at rate $\lambda $. Let $n_3$ (respectively, $n_4$ and $n_5$) denote the number of links between nodes of type $S$ and $I$ (respectively, $S$ and $R$ and $R$ and $I$).
In the infinite population limit, with the assumptions of spatial homogeneity and uncorrelated pairs, the system is described by the deterministic equations of the standard or uncorrelated PA as follows {\cite{lebowitz}}: 
\begin{eqnarray}
\label{paeq} \dot{p}_1&=&\gamma(1-p_2-p_1)-k\lambda p_3 \ , \\
\dot{p}_2&=&k\lambda p_3-\delta p_2 \ , \nonumber\\
\dot{p}_3&=&\gamma p_5-(\lambda+\delta)p_3 + \frac{(k-1)\lambda p_3}{p_1}(p_1-p_4-2p_3) \ , \nonumber\\
\dot{p}_4&=&\delta p_3+\gamma(1-p_1-p_2-p_5-2p_4)-\frac{(k-1)\lambda p_3p_4}{p_1} \ , \nonumber\\
\dot{p}_5&=&\delta(p_2-p_3)-(\gamma+2\delta)p_5+\frac{(k-1)\lambda p_3p_4}{p_1} \ . \nonumber
\end{eqnarray}
In the above equations the variables stand for the limit values of the 
node and pair densities $p_1=n_1/N$, $p_2=n_2/N$, $p_3=n_3/(kN)$, $p_4=n_4/(kN)$, and $p_5=n_5/(kN)$ as $N \to \infty $.
As expected, neglecting the pair correlations and setting $p_3=p_1p_2$ in the first two equations leads to the classic equations of the randomly mixed or mean-field approximation (MFA) SIRS model,
\begin{eqnarray}
\label{mfeq}\dot{p}_1&=&\gamma(1-p_2-p_1)- k \lambda p_1p_2 \ , \\
\ \dot{p}_2&=&k \lambda p_1p_2-\delta p_2 \ . \nonumber
\end{eqnarray}
The phase diagrams of the two models are plotted in
Fig. \ref{phasediagramSIRS} [solid lines for Eq. (\ref{paeq}) and dashed line for Eq. (\ref{mfeq}), both with $k=4$]. We have set the time scale so that $\delta=1$.
The critical line separating a susceptible-absorbing phase from an active phase where a stable steady state exists with nonzero infective density is given by ${\lambda}_{c}^{\text{MFA}}=\frac{1}{4}$ (dashed black line) for the MFA, and by $\lambda_{c}^{\text{PA}}(\gamma)=\frac{\gamma +1}{3\gamma+2}$ [solid orange (gray) line]
for the PA. In addition, in the active phase of the PA we find for small values of $\gamma $ a new phase boundary
[solid blue (black) line] that corresponds to a Hopf bifurcation and seems
to have been missed in previous studies of this model {\cite{lebowitz}}. This boundary
separates the active phase with constant densities (region I) from an active phase with oscillatory
behavior (region II).  The maximum of the curve is situated at
$\lambda \approx 2.5$, $\gamma \approx 0.03$, which means that the PA model predicts sustained oscillations in the thermodynamic limit when loss of immunity is much slower than recovery from infection. According to published data for childhood infections in
the pre-vaccination period \cite{bauschearn}, 
taking the average immunity waning time to be of the order of the length of the
elementary education cycles at that time   
(10 years for the data points in Fig. \ref{phasediagramSIRS}) many of the estimated parameter values for these diseases fall into oscillatory region II, and the others are in region I close to the boundary. Different data points for the same disease correspond to estimates for $\lambda$ based on different data records. Although small enough to be missed in a coarse grained numerical
study, the oscillatory phase is large in the admissible parameter region of an important class of diseases. A systematic study of the dependence of this oscillatory phase on the parameter $k$ and of its relevance to understand the behaviour of simulations on networks will be reported elsewhere {\cite{sirsnetworks}}. Preliminary results indicate that the oscillatory phase persists 
in the range $2< k \lesssim 6$, and that it gets thinner as $k$ increases. There are indications that this oscillatory phase is robust also with respect to variations of the underlying model
\cite{robust}.

Let us now study the combined effect of correlations and demographic stochasticity in region I by
taking $N$ large but finite. In the stochastic version of the MFA SIRS model the state of the
system is defined by  $n_1$ and $n_2$ which change according to the transition rates as
\begin{eqnarray}
\label{ratesmfa}
\label{ratesmf}T^{n_1+1,n_2} &=& \gamma \ (N-n_1-n_2) \ , \\
T^{n_1,n_2-1} &=& \delta \ n_2 \ , \nonumber\\
 T^{n_1-1,n_2+1} &=& \ k\lambda \ \frac{n_1}{N} \ n_2 \ , \nonumber
\end{eqnarray}
associated to the processes of immunity waning, recovery and infection. Here $T^{n_1+k_1,n_2+k_2}$
denotes the transition rate from state $(n_1, n_2)$ to state $(n_1+k_1,n_2+k_2)$, $k_i \in \{-1,0,1 \}$, where $i=1,2$.
As in \cite{McKane-Newman2005}, the power spectrum of the normalized fluctuations (PSNF)
around the active steady state of
system (\ref{mfeq}) can be computed approximately from the next-to-leading-order terms of
van Kampen's system size expansion of the corresponding master equation. Setting $n_1(t)=
N p_1(t) + \sqrt{N} x_1(t)$ and $n_2(t)= N p_2(t) + \sqrt{N} x_2(t)$, the equations of motion for the average densities (\ref{mfeq}) are given by the leading-order terms of the expansion. The next-to-leading-order terms yield a linear Fokker-Planck equation for the probability distribution function 
$\Pi(x_1(t),x_2(t),t)$. The equivalent Langevin equation for the normalized fluctuations is 
$\dot {x}_i (t) = \sum_{j=1}^2{J}_{ij} {x}_j(t) + {L}_i(t)$, where ${\bf J}$ is the Jacobian of Eq. (\ref{mfeq}) at the endemic equilibrium
and ${L}_i(t)$ are Gaussian white noise terms whose amplitudes are
given by the expansion. Taking the Fourier transform we obtain for the PSNF,
\begin{equation}
\label{psgen}
P_i(\omega)\equiv\langle |\tilde x_i(\omega)|^2 \rangle  = \sum\limits_{j,k}M_{ik}^{-1}(\omega)B_{kj}M_{ji}^{-1}(-\omega) \ ,
\end{equation}
where $M_{ik}(\omega)=\text{i} \omega \delta_{ik} - J_{ik}$ and $\langle \tilde L_i(\omega) \tilde L_j (\omega') \rangle = B_{ij} \delta (\omega+\omega')$. For $k=4$ and $\delta =1$ this expression becomes
\begin{eqnarray}
P_{\text{S}}^{\text{MFA}}&=&\frac{B_{11}(J_{12}^2+{\omega}^2)}{(D-{\omega}^2)^2+T^2{\omega}^2} \ , \\
P_{\text{I}}^{\text{MFA}}&=&\frac{B_{11} (J_{11}^2+J_{11}J_{21}+J_{21}^2+{\omega}^2)}{(D-{\omega}^2)^2+T^2{\omega}^2} \ ,
\end{eqnarray}
\noindent{}where $D$ and $T$ are the determinant and the trace of ${\bf J}$ and $B_{11}=B_{22}=-2B_{12}=-2B_{21}=\frac{\gamma (4\lambda -1)}{2\lambda (\gamma+1)}$, for the susceptible and the infected PSNFs, respectively.

In a stochastic version of the PA SIRS model the state of the system is defined by the integers $n_i$, where $i=1, ...,5$, and recovery, loss of immunity, and infection induce different transitions according to the pairs or triplets involved in the process. The simplest set of transitions and transition rates compatible with Eq. (\ref{paeq}) is
\begin{eqnarray}
\label{ratespa}
&&T^{n_1+1,n_2,n_3+k,n_4,n_5-k}=\frac{\gamma}{k}n_5 \ , \\
&&T^{n_1+1,n_2,n_3,n_4-k,n_5}=\frac{\gamma}{k}n_4 \ , \nonumber \\
&&T^{n_1+1,n_2,n_3,n_4+k,n_5}=\frac{\gamma}{k}(k(N-n_1-n_2)-n_4-n_5) \ , \nonumber \\
&&T^{n_1,n_2-1,n_3-k,n_4+k,n_5}=\frac{\delta}{k}n_3 \ , \nonumber \\
&&T^{n_1,n_2-1,n_3,n_4,n_5+k}=\frac{\delta}{k}(kn_2-n_3-n_5) \ , \nonumber \\
&&T^{n_1,n_2-1,n_3,n_4,n_5-k}=\frac{\delta}{k}n_5 \ , \nonumber \\
&&T^{n_1-1,n_2+1,n_3-k,n_4,n_5}=\frac{\lambda}{k}\frac{n_3}{n_1}n_3 \ , \nonumber \\
&&T^{n_1-1,n_2+1,n_3-1,n_4-(k-1),n_5+(k-1)}=\frac{\lambda}{k}\frac{n_3}{n_1}n_4 \ , \nonumber\\
&&T^{n_1-1,n_2+1,n_3+(k-2),n_4,n_5}=\frac{\lambda}{k}\frac{n_3}{n_1}(kn_1-n_3-n_4) \ .\nonumber
\end{eqnarray}
This is a coarse grained description where the effect of the change in state of a given node on the
$k$ pairs that it forms is averaged over each pair type. For instance, the event of loss of immunity occurs at a rate $\gamma n_R$, where $n_R$ is the number of recovered nodes, and changes the $k$ pairs formed by the node that
switches from recovered to susceptible. On average, each pair type will change by $k$ units at a rate proportional to its density, according to the equation $\gamma n_R = \gamma n_R(\frac{n_4}{kn_R}+\frac{n_5}{kn_R}+\frac{2n_{RR}}{kn_R})$,
where $n_{RR}$ is the number of pairs of recovered nearest neighbours. Taking this level of description and using $kn_R=n_4+n_5+2n_{RR}$ and $n_1+n_2+n_R=N$, we obtain the first three equations of Eq. (\ref{ratespa}) for the rates of the three different pair events associated with loss of immunity. A full microscopic description would require considering separately 
all possible five-node configurations for the central node that switches from $R$ to $S$ and its four nearest neighbours. We have checked that the detailed stochastic model involving 40 different transitions for $k=4$ gives essentially the same results {\cite{sirsnetworks}} as the coarse grained model (\ref{ratespa}) that we consider here.

For the fluctuations of the pair densities we set $n_3(t)=Nk p_3(t) + \sqrt{N}k x_3(t)$, $n_4(t)=Nk p_4(t) + \sqrt{N}k x_4(t)$, and $n_5(t)=Nk p_5(t) +\sqrt{N}k x_5(t)$. 
The leading order terms of van Kampen's system size expansion of the master equation associated to
(\ref{ratespa}) yield the deterministic PA Eqs. (\ref{paeq}). An approximate analytical expression for the PSNF can be obtained as before from the next-to-leading-order terms.
Formula (\ref{psgen}) is still valid taking now ${\bf J}$ as the Jacobian of Eq. (\ref{paeq}) at the endemic equilibrium and the noise cross correlation matrix ${\bf B}$ computed directly from the expansion. 
\begin{figure}[h]
{\includegraphics[width=\columnwidth]{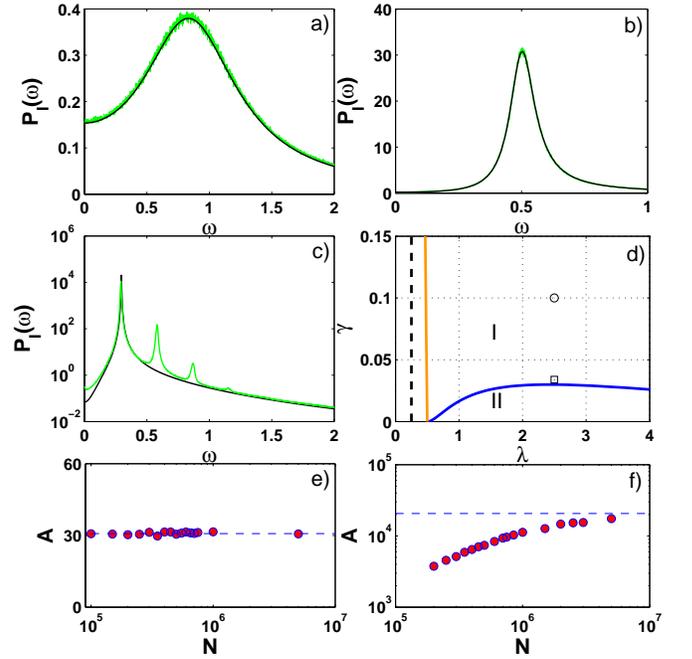}
\caption{(Color online) (a) 
Analytical and numerical PSNFs of the infectives for model (\ref{ratesmf})
with $\gamma=0.1$ and $\lambda=2.5$; (b) the same for model (\ref{ratespa});
(c) a similar plot for model (\ref{ratespa}) with $\gamma=0.034$ and $\lambda=2.5$, notice the 
lin-log scale; 
(d) location of the parameter values chosen for
 (a) and (b) (circle) and for (c) (square); and
(e) and (f) plots of the peak amplitude of the PSNF of the PA model as a function of $N$ for the 
parameter values chosen for (b) and (c).
}\label{figure2}}
\end{figure} 

In Fig. \ref{figure2} the approximate PSNFs given by Eq. (\ref{psgen})
are plotted (black lines) and compared with the numerical power spectra of stochastic simulations for $N=10^6$ [green (gray) lines] for models (\ref{ratesmf}) and (\ref{ratespa}) [Figs. \ref{figure2}(a) and \ref{figure2}(b)].
For this system size, there is almost perfect agreement between the analytical
approximate expressions and the results of the simulations across the whole region I.
The plots show that for the same parameter values the fluctuations are larger and more
coherent for model (\ref{ratespa}), in agreement with the results of simulations reported in
\cite{interface} for small-world networks on a lattice and variable small-world parameter.
This effect becomes much more pronounced as the boundary between regions I and II, where the analytical PSNF of model (\ref{ratespa}) diverges, is approached.
Close to this boundary [see Fig. \ref{figure2}(c)], there is a significant discrepancy between the analytical (black line) and the numerical [green (gray) line] PSNFs associated with the appearance of secondary peaks at multiples of the main peak frequency. This is a precursor of the oscillatory phase, and the breakdown of van Kampen's approximation for this system size may be understood as an effect of the loss of stability of the endemic equilibrium close to the boundary. Relaxation towards equilibrium
becomes slow compared with the period of the damped oscillations, and a significant part
of the power spectrum energy shows up in the secondary harmonics. For these parameter values, van Kampen's expansion becomes a good approximation only for larger system sizes. 
Also shown in Fig. \ref{figure2}(e) [respectively, Fig. \ref{figure2}(f)] is
the scaling with system size of the peak amplitude of the infectives PSNF of the PA model [pink (black) dots] for the parameter values considered in (b) [respectively, (c)] and the peak amplitude (dashed blue line) of the approximated PSNF given by Eq. (\ref{psgen}).
Away from the phase boundary of the oscillatory phase we find that the simulations exhibit the amplitude and scaling predicted by Eq. (\ref{psgen}) down to system sizes of $\sim 10^5$. By contrast, close to the phase boundary the match is reached only for system sizes larger than $5\times10^6$. 
\begin{figure}[h]
{\includegraphics[width=\columnwidth]{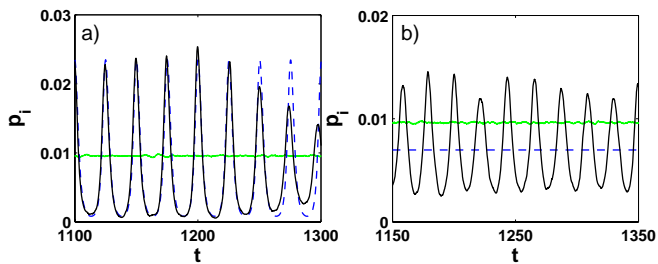} \caption{(Color online) The steady-state infective density given by the PA deterministic model (dashed blue 
lines), and by simulations of the PA and the MFA stochastic models [solid black and green (gray) lines, respectively] in regions II and I for $N=10^7$. Parameters are (a) $\lambda=7.5$, $\gamma=0.01$ and (b) $\lambda=9$, $\gamma=0.01$.}
\label{figure3} }
\end{figure}

Examples of typical time series predicted by the PA model in the parameter region
of childhood infectious diseases are shown in Fig. \ref{figure3} [dashed blue lines for the 
deterministic model (\ref{paeq}) and solid black lines for simulations of the stochastic model (\ref{ratespa})]. The results of simulations of the MFA stochastic model (\ref{ratesmfa}) for the same parameter values are also shown for comparison [solid green (gray) lines]. 
Fig. \ref{figure3}(a) illustrates the regular high-amplitude oscillations of region II. 
All over this region, simulations of the stochastic model (\ref{ratespa}) reproduce
the behaviour of the solutions of Eq. (\ref{paeq}) with added amplitude fluctuations.
The only limitation to observe these oscillations in finite systems is that $N$ has to be
taken large enough for the deep interepidemic troughs to be spanned without stochastic extinction
of the disease. In region I [Fig. \ref{figure3}(b)] there are no oscillations in the thermodynamic
limit but, in contrast to the stochastic MFA model, the resonant fluctuations in the PA model are large and coherent enough to provide a distinct cycling pattern, which is partially described by van Kampen's expansion (\ref{psgen}).
 
In conclusion, we have considered a stochastic version of the basic model of infection
dynamics including a representation of the spatial correlations of an interaction network
through the standard PA. We have shown that in general the resonant amplification 
and the coherence of stochastic fluctuations are much enhanced with respect to the MFA model.
This quantitative difference becomes qualitative in a region of parameter space
that corresponds to diseases for which immunity waning is much slower than recovery. 
In this region
the nonlinearities of the model and demographic stochasticity give rise either to oscillations 
that persist in the thermodynamic limit or to high amplitude, coherent resonant fluctuations,  providing realistic 
patterns of recurrent epidemics. 

These results are relevant for other population dynamics models in the slow driving regime that corresponds to small $\gamma $ in our model, 
suggesting that in systems of moderate size intrinsic stochasticity together
with the simplest representation of spatial correlations may be enough to produce distinct
oscillatory patterns.  
This favours the view that, for a large class of systems, noisy oscillations in population dynamics data may be intrinsic, rather than externally driven. 

Financial support from the Foundation of the University of Lisbon 
and the Portuguese Foundation for Science and 
Technology (FCT) under Contracts No. POCI/FIS/55592/2004
and No. POCTI/ISFL/2/618 is gratefully 
acknowledged. The first author (G.R.) was also supported by FCT under Grant No. SFRH/BD/32164/2006 and by Calouste Gulbenkian Foundation under its Program "Stimulus for Research".


\begin{thebibliography}{99}
\section{REFERENCES}
\bibitem{epid_osc}
P.~Rohani, D.~J.~D.~Earn, and B.~T.~Grenfell, 
Science {\bf 286}, 968 (1999); N.~C.~Grassly, C.~Fraser, and G.~P.~Garnett,
Nature (London) {\bf 433}, 417 (2005).

\bibitem{ref_contro}
O.~N.~Bj\o rnstad and B.~T.~Grenfell,
Science {\bf 293}, 638 (2001).

\bibitem{seasonal}
M.~J.~Keeling, P.~Rohani, and B.~T.~Grenfell, 
Physica D {\bf 148}, 317 (2001).

\bibitem{bauschearn}
C.~T.~Bauch and D.~J.~D.~Earn, 
Proc. R. Soc. London, Ser. B {\bf 270}, 1573 (2003).

\bibitem{mathlit}
H.~W.~Hethcote and P.~van~den~Driessche,
J. Math. Biol. {\bf 29}, 271 (1991);
W.~Wang, 
Math. Biosci. Eng. {\bf 3}, 267 (2006).

\bibitem{grenfell_measles}
D.~J.~D.~Earn, P.~Rohani, B.~M.~Bolker, and B.~T.~Grenfell, 
Science {\bf 287}, 667 (2000);
B.~T.~Grenfell, O.~N.~Bj\o rnstad, and J.~Kappey,
Nature (London) {\bf 414}, 716 (2001).

\bibitem{bausch}
C.~T.~Bausch, in \textit{Mathematical Epidemiology}, edited by F.~Brauer, P.~van~den~Driessche, and J.~Wu (Springer, Berlin, 2008), p. 297.

\bibitem{physics_sims}
J.~E.~Satulovsky and T.~Tom{\'e}, Phys. Rev. E {\bf 49}, 5073 (1994);
A.~Lipowski, Phys. Rev. E {\bf 60}, 5179 (1999); 
M.~Kuperman and G.~Abramson,
Phys. Rev. Lett. {\bf 86}, 2909 (2001);
T.~Gross, Carlos~J.~Dommar~DLima, and B.~Blasius,
Phys. Rev. Lett. {\bf 96}, 208701 (2006).

\bibitem{McKane-Newman2005}
A.~J.~McKane and T.~J.~Newman, 
Phys. Rev. Lett. {\bf 94}, 218102 (2005).

\bibitem{McKane_inter}
D.~Alonso, A.~J.~McKane, and M.~Pascual, 
J. R. Soc., Interface {\bf 4}, 575 (2007).


\bibitem{bartlett}
M.~S.~Bartlett, \textit{Stochastic Population Models in Ecology and Epidemiology} (Methuen, 
London, 1960).


\bibitem{vanKampen}
N.~G.~van~Kampen, \textit{Stochastic Processes in Physics and Chemistry} (Elsevier, Amsterdam, 1981).

\bibitem{lebowitz}
J.~Joo and J.~L.~Lebowitz, Phys. Rev. E {\bf 70}, 036114 (2004).

\bibitem{rev_keeling}
M.~J.~Keeling and K.~T.~D.~Eames, 
J. R. Soc., Interface {\bf 2}, 295 (2005).

\bibitem{interface}
M.~Sim\~oes, M.~M.~Telo~da~Gama, and A.~Nunes,
J. R. Soc., Interface {\bf 5}, 555 (2008).


\bibitem{sirsnetworks}
G.~Rozhnova and A.~Nunes, e-print arXiv:0812.1812.


\bibitem{robust}
D.~A.~Rand, in \textit{Advanced Ecological Theory: Principles and Applications}, edited by J.~McGlade (Blackwell Science, Oxford, 1999), p. 100; J.~Benoit, A.~Nunes, and M.~M.~Telo~da~Gama, Eur. Phys. J. B {\bf 50}, 177 (2006).

\end{thebibliography}
\end{document}